# Stochastic Memristive Interface between Electronic FitzHugh-Nagumo Neurons

**Svetlana Gerasimova[1,*], Alexey Belov[1], Dmitry Korolev[1], Davud Guseinov[1], Albina Lebedeva[1], Maria Koryazhkina[1], Alexey Mikhaylov[1], Victor Kazantsev[1,2,3], Alexander N. Pisarchik[1,3,4]**

[1]National Research Lobachevsky State University of Nizhny Novgorod, Nizhny Novgorod, Russia

[2]Innopolis University, Innopolis, Russia

[3]Immanuel Kant Baltic Federal University, Kaliningrad, Russia

[4]Center for Biomedical Technology, Universidad Politécnica de Madrid, Pozuelo de Alarcón, Madrid, Spain

**\* Correspondence:**
Svetlana A. Gerasimova
gerasimova@neuro.nnov.ru

## Abstract

The dynamics of memristive device in response to neuron-like signals and coupling electronic neurons via memristive device has been investigated theoretically and experimentally. The simplest experimental system consists of electronic circuit based on the FitzHugh-Nagumo model and metal-oxide memristive device. The hardware-software complex based on commercial data acquisition system is implemented for the imitation of signal from presynaptic neuron`s membrane and synaptic signal transmission between neurons. The main advantage of our system is that it uses real time dynamics of memristive device. Electrical response of memristive device shows its behavioral flexibility that allows presenting a memristive device as an active synapse. This means an internal adjustment of the parameters of memristive device that leads to modulation of neuron-like signals. Physics-based dynamical model of memristor is developed in MATLAB for numerical simulation of such a memristive interface to describe and predict experimentally observed regularities of synchronization of neuron-like oscillators. FitzHugh-Nagumo circuits time series with a linear or stepwise increase in the signal amplitude are used to study the memristor response and coupling of neuron-like oscillators taking into account the stochasticity of memristor model to compare the numerical and experimental data. The observed forced synchronization modes characterize the dynamic complexity of the memristive device, which requires further description using high-order dynamical models. The developed memristive interface will provide high efficiency in the imitation of the synaptic connection due to its stochastic nature and can be used to increase the flexibility of neuronal connections for neuroprosthetic challenges.

## INTRODUCTION

The design of compact neuromorphic systems (including micro- and nanochips) capable of reproducing the information and computing functions of brain cells is a great challenge of the modern science and technology. Such systems are of interest for both fundamental research in the field of the nonlinear dynamics of complex and multistable systems (Alombah, 2017, García-Vellisca, 2017) and medical application in devices for monitoring and stimulating the brain activity in the framework of neuroprosthetic tasks (Horch, 2004). In this context, memristive devices have been an object of intensive research in recent years including neuromorphic applications (Indiveri, 2011, Kuzum, 2013, Bill, 2014, Zhang, 2017, Nair, 2017, Strukov, 2018, Mikhaylov, 2020). It is worth noting that the construction and creation of electronic neurons and synapse (connections between neurons) based on thin-film memristive nanostructures is one of the most rapidly developing area of interdisciplinary research in the development of neuromorphic systems (Thomas, 2013, Adamatzky, 2014, Ge, 2017). Neuromorphic technologies are relevant to intellectual adaptive automatic control systems, biorobots.

The history of neuromorphic technologies began in the late 1980s, and significant advances have been achieved ever since thanks to the progress in electronics, physics of micro- and nanostructures, and solid state nanoelectronics. Widely studied neuron-like electric circuits allow reproducing both qualitative main neuron characteristic and complex dynamic regimes of neurons (Binczak, 2006, Shchapin, 2009, D. Adamchik, 2015, Gerasimova, 2015, Mishchenko, 2017).

Memristive device represents a physical model of a Chua`s memristor (A. Adamatzky, 2014), which is an element of electric circuits capable of changing the resistance depending on the electric signal received at the input. In the last decades, different thin-film memristive nanostructures have been designed to generate neural dynamics (Gambuzza, 2017, Guseinov, 2017). Moreover, the ability of a memristive structure to change conductivity under the action of pulsed signals makes it an almost ideal electronic analogue of a synapse (Bill, 2014). A synapse as known to represent a communication channel between neurons. Neurons interact through synapses, which ensure the unidirectional transmission of signals from a transmitting (presynaptic) neuron to a receiving (postsynaptic) one. The communication channel is ensured by propagation of a neural impulse along the axon of the transmitting cell. One of the effects of synaptic coupling is forced synchronization of the receiving neuron with the transmitting one. Forced synchronization was observed in models of various physical phenomena and can often be described in terms of periodic solutions of dynamical systems (Matrosov, 2011, Matrosov 2013, Selyutskiy, 2017, Boccaletti, 2018). The synaptic electronic circuits were built to transform presynaptic voltage pulses to postsynaptic currents with some synaptic gains. Different strategies of hardware implementation of synaptic circuits were employed, for example, the optical interface between electronic neuron models (Pisarchik 2011, Pisarchik, 2013, Gerasimova, 2015).

Recent advances in nanotechnologies allowed the miniaturization of artificial synapses by constructing memristive nanostructures which mimic synaptic dynamics. In particular, among various candidates for electronic synapses, memristive synaptic devices have the highest potential to realize massive parallelism and 3D integration for achieving good functionality per unit volume (Matveyev, 2015, Shi 2017, Choi, 2018). The current challenge is the integration of the neuron-like signals (spikes) into a memristor-based neuromorphic system. The interaction between such artificial neurons via metal-oxide memristive device was successfully implemented in hardware (Gerasimova, 2017). A prerequisite for such work was the study of the interaction of Van der Pol generators through a memristor (Ignatov, 2015). To date, a significant amount of theoretical research has been performed on the study of synchronization between neuron-like generators connected through a memristive device (Zhang, 2017, Korotkov, 2019). However, to our knowledge there are no experimental studies on the dynamics of the FitzHugh-Nagumo circuits memristor-coupled systems.

The development of neuro-hybrid systems should lead to the creation of simple and compact neuroelements based on memristive devices and capable to mimic electrophysiological behavior of real neurons. In this work, we simulate and experimentally implement a memristive interface of coupled electronic FitzHugh-Nagumo generators on the basis of metal-oxide nanostructure. This interface allows to simulate and predict the effects of adaptive behavior and synchronization of neurons. We also investigate stochastic nature of the memristive device and describe system parameters in real time.

## MATHEMATICAL MODELING OF MEMRISTIVE INTERFACE

To simulate the dynamics of an individual neuron, we used the FitzHugh-Nagumo generator with cubic nonlinearity performed using diodes (Shchapin, 2009, Gerasimova, 2015). The FitzHugh-Nagumo system of differential equations is built by normalizing the equations obtained in accordance with the Kirchhoff law (Binczak, 2006). This model contains a potential-like variable ($u$) and a "recovery function" that determines the dynamics of the ion current ($v$), cubic nonlinearity, a depolarization parameter characterizing the excitation threshold ($I$), and a small parameter of the system $\varepsilon$.

The FitzHugh-Nagumo model is given by the following equations

$$\frac{du}{dt} = f(u) - v$$
$$\frac{dv}{dt} = \varepsilon(g(u) - v) - I,$$

where $u$ is the membrane potential of presynaptic neuron, $v$ is the "recovery" variable, $f(u) = u - u^3/3$ is cubic nonlinearity, $g(u) = \alpha u$, if $u < 0$, and $g(u) = \beta u$, if $u \geq 0$, $\varepsilon$ is a small parameter, and $\alpha$, $\beta$ control respectively the shape and location of the $v$-nullcline (Binczak, 2006). This system of ordinary differential equations is simulated by the Runge-Kutta method in Matlab.

A memristive device model is designed based on the standard approach to reflect the dynamic response of the memristor to the electrical stimulation. The model describes an analog change in resistance, similar to potentiation and depression, based on the physical laws revealed in experiments (Gerasimova, 2018).

The model of a memristive device can be expressed as

$$j = wj_{lin} + (1 - w)j_{nonlin}$$
$$j_{lin} = u/\rho$$
$$j_{nonlin} = u\exp(b\sqrt{u} - E_b)$$
$$w(t, u) = A\exp(-\frac{E_m - \alpha_1 u}{kT})$$

Our approach involves the introduction of the internal state parameter $w$ determined by the fraction of the area of the structure occupied by conducting filaments and its change due to migration of oxygen vacancies (effective barrier for migration $E_m$) activated by Joule heating ($kT$) and electric field / voltage ($u$). The full current density ($j$) through the memristive structure consists of the linear component ($j_{lin}$) which corresponds to the ohmic conductivity ($\rho$ is the resistivity) through the filaments and nonlinear ($j_{nonlin}$) determined by the charge carriers transport through defect states in the oxide material in the region of filament rupture or in the rest of the structure. The nonlinear transport of charge carriers (an effective barrier $E_b$) is described by the Frenkel-Poole law, based on the approximation of the current-voltage characteristics in the high resistance state (Gerasimova, 2018). A smooth transition between high and low-resistance states is determined by the dynamic contribution to the total current of conductive filaments, and therefore the state parameter. In these equations t is the time $b$, $\alpha_1$ and $A$ are coefficients determined from the experimental data.

To compare the experimental data for memristive devices with the simulation results, the mathematical model of the memristive device takes into account the stochasticity of microscopic phenomena that lead to a change in the internal state $w$ of the dynamical system. A random variable fluctuation according to the normal law is added to the model parameters that allows obtaining a scatter of the current-voltage characteristics of the experimental curves. The model considers three main sources of stochasticity: the energy barrier for ion hopping $E_m$ (dispersion 10%), the energy barrier $E_b$ for electron jumps in the Poole-Frenkel conduction mechanism in the high-resistance state (dispersion 1%), and the ohmic resistance of the structure in the conducting state $\rho$ (dispersion 10%). The final scatter in switching voltages is mainly related to the stochasticity of the energy barrier for ions, while the change of resistive states from cycle to cycle is associated with the stochasticity of electron transport.

The unilaterally coupling between model neurons via the memristive device leading to a master-slave configuration also is modeled by the following equations

$$\frac{du_i}{dt} = f(u_i) - v_i + j(u_1)d\delta_{2,i}$$

$$\frac{dv_i}{dt} = \varepsilon(g(u_i) - v_i) - I_i \quad ,$$

where $i = \{1,2\}$, $d$ is equivalent load resistance, $\delta_{2,i}$ is Kronecker symbol, so that $\delta_{2,1} = 0$, $\delta_{2,2} = 1$, $j(u_1)$ is the current through memristive device.

Therefore, the two model neurons are coupled so that a part of current weighted by memristive device current $j(u_1)$ via load resistance and generated by first neuron (master) is included in the second neuron (slave). The initial conditions and parameters of the model correspond to the experiments. Specifically, at the initial moment of time, the Master and Slave oscillators are in the self-oscillatory regime.

## ELECTRICAL MEMRISTIVE INTERFACE

The designed neuromorphic circuit consists of electronic circuit based on the FitzHugh-Nagumo model, thin-film metal–oxide–metal nanostructures based on yttria-stabilized zirconium oxide (Au/ZrO$_2$(Y)/TiN/Ti) as a memristive device (Emelyanov, 2019) and current compliance. This memristive interface operates as follows. The electronic neuron based on the FitzHugh-Nagumo circuit generates pulse signal acting on the memristive device and thus modulating the oxidation and recovery of segments of conducting channels (filaments) in the memristive device oxide film.

It should be noted that the FitzHugh-Nagumo neuron dynamics qualitatively exhibits the main features of a living neuron, namely, the presence of an excitability threshold, and resting and spiking behaviors. The potentiometer in the circuit provides the control over dynamical modes of the FitzHugh-Nagumo generator, i.e. its resting and spiking states. The spiking frequency of the FitzHugh-Nagumo neuron is varied in the range of 10 – 150 Hz, the spike duration in the range of 10 – 25 ms, and the spike amplitude in the range of 1 – 6 V. Note that the current is limited in the negative voltage area using the field transistor 2N3329 and diode 1N913 (see Figure 1).

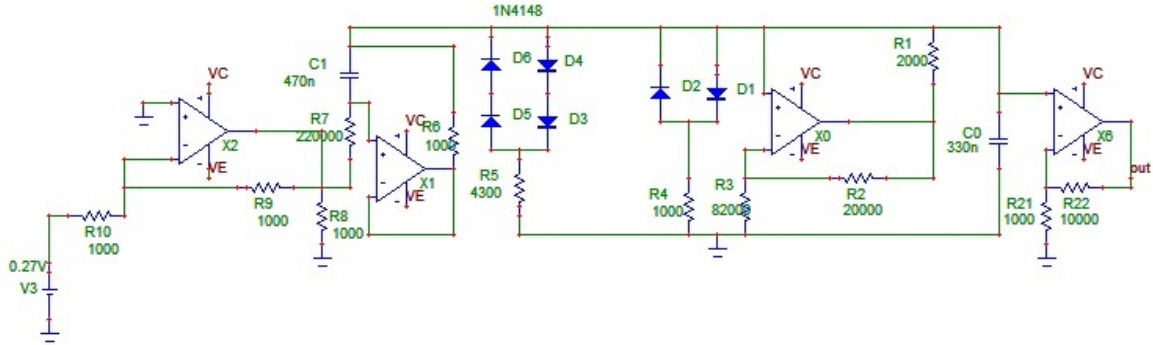

**Figure 1. Electronic scheme of analog electrical FitzHugh-Nagumo neuron.** The inductance is implemented by the circuit with the operational amplifier. Cubic nonlinearity is obtained using the set of diodes, capacitor C0 simulates the capacitance of the neuron membrane and potential V3 simulates the equilibrium potential controlled by a power source.

The hardware-software complex is based on (DAQ) National Instruments USB-6212, consisting of one a digital-to-analog converter (DAC) and two analog-to-digital converters (ADC). The DAC cyclically reproduces a pre-recorded neuron-like signal with a sampling frequency of 5 kHz. The current via the memristive device is calculated using the load resistance potential difference. Each of ADC separately registers the voltage drop on the memristive device and the load resistor, and allows the calculation of the memristive device resistance in real time independently of other circuit elements. The software is developed by the National Instruments LabVIEW. The potential difference on the memristive device ($R_m$) and the load resistance ($R_2$) is digitized with a sampling frequency of 10 kHz. Experimental data are analyzed using the MatLab software.

After testing and tuning, the neuron-like oscillators were connected through the memristive device. The experiment was carried out as follows. Master and slave analog neurons were turned in oscillatory regimes. The signal from the master generator was sent to the slave generator through the memristive device. Under the neuron-like signal action, the memristive device changed its state from low resistive to high resistive. The amplitude of the master analog neuron was varied by the potentiometer so that the signals from oscillators were frequency-locked.

## SWITCHING OF MEMRISTIVE DEVICE UNDER THE NEURON-LIKE SIGNAL ACTION

The neuron-like signal affecting the memristive device is shown in Figure 2 (A). Note the asymmetry of the neuron-like signal: the minimum peak amplitude is –5 V, and the maximum peak is 4 V. Each interval is highlighted in color corresponding to the interval of the current-voltage characteristic (*I-V* curves) of Figure 2(B). This allows tracing the effect of the neuron-like signal in more detail. The *I-V* curves of Figure 2(B) demonstrate the switching between the low-resistance state (LRS) and the high-resistance state (HRS). The corresponding current changes are indicated in Figure 2(B) as RESET and SET. Figure 2(B) shows the scatter of the *I-V* curves obtained by adding a random value to the parameters, i.e. the energy barrier for the migration of ions $E_m$, the energy barrier for hopping of electrons $E_b$, and the ohmic resistance of the device in the conducting state. Figure 2(C) demonstrates increasing the signal amplitude from 1.558 V to 4 V. Figure 2(D) illustrates the stochastic effect of switching on the corresponding *I-V* curves.

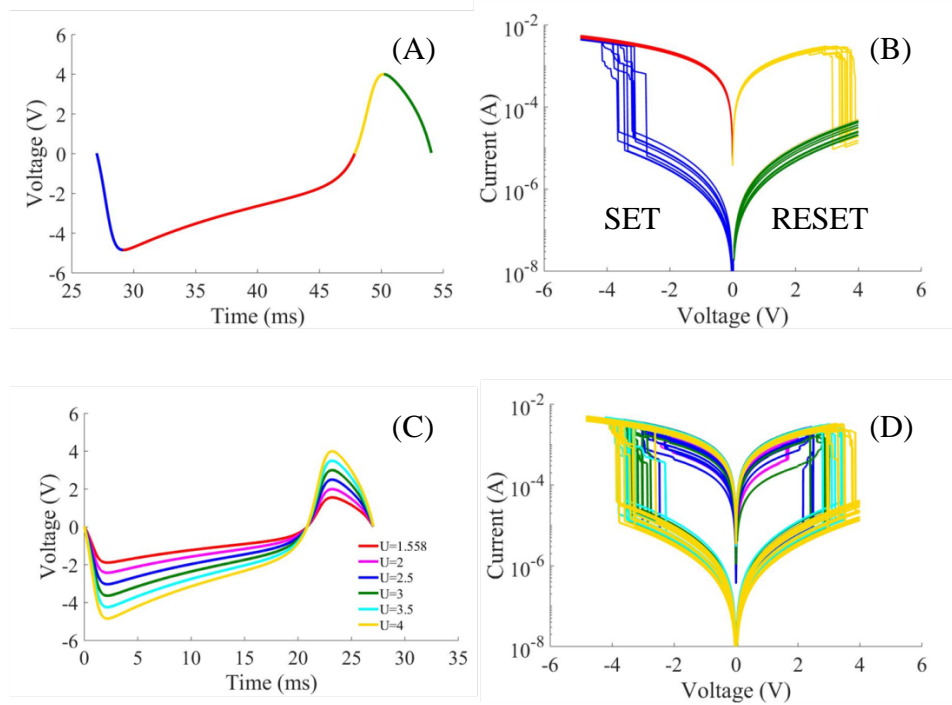

**Figure 2. The resistive switching in response to a neuron-like signal.** (A) Neural-like pulse. (B) *I-V* curves. (C) Increasing the amplitude of neuron-like signal. (D) Resistive switching of memristive device under the action of corresponding neuron-like signals on *I-V* curves.

In Figure 2(C) one can see that even when exposed to a small amplitude signal of 1.558 V (red curve), there is a possibility of switching the memristive device from a state with a high resistance to a state with a low resistance. As seen from Figure 2(D), for curves (yellow color) with amplitude of 4 V, a stable reproducible switching is achieved which corresponds to experiments.

## MEMRISTIVE COUPLING BETWEEN MODEL AND ANALOG NEURONS

We found that with an increase in the amplitude of a neuron-like signal, starting from a certain threshold value, stochastic switching of the memristor is observed at each spike, as well as the general evolution of the state of the memristor. At high oscillation amplitudes, the system goes into an extreme resistive state and stops responding to each spike or switches between extreme states. In the course of the study, the optimal coupling strength was established for synchronization 1:1, $d = j$ $(u_1)$ $R = (0.02 - 0.06)$, for fractional synchronization $d = (0.06 - 0.095)$ (see Figure 3).

A series of experiments were carried out to show that at the amplitude of the master generator from 1.6 to 2 V, the synchronization modes of the master and slave generators 1:1, 2:1 are switched for a certain time $(1 - 3 \text{ s})$ (see Figure 4). It was also shown that the memristive coupling between FitzHugh-Nagumo neurons leads to synchronization regimes and intermittency mode to chaos. It was found that spontaneous switching between synchronization modes is possible. Thus, the experimental behavior includes simple synchronization modes observed in simulation, as well as a more complex dynamics which cannot be described by a first-order memristor model, even if the stochasticity of switching is taken into account. Therefore, higher-order memristor models based on two or more state variables should be used to implement complex intermittency and chaotic modes of synchronization (Guseinov, 2021).

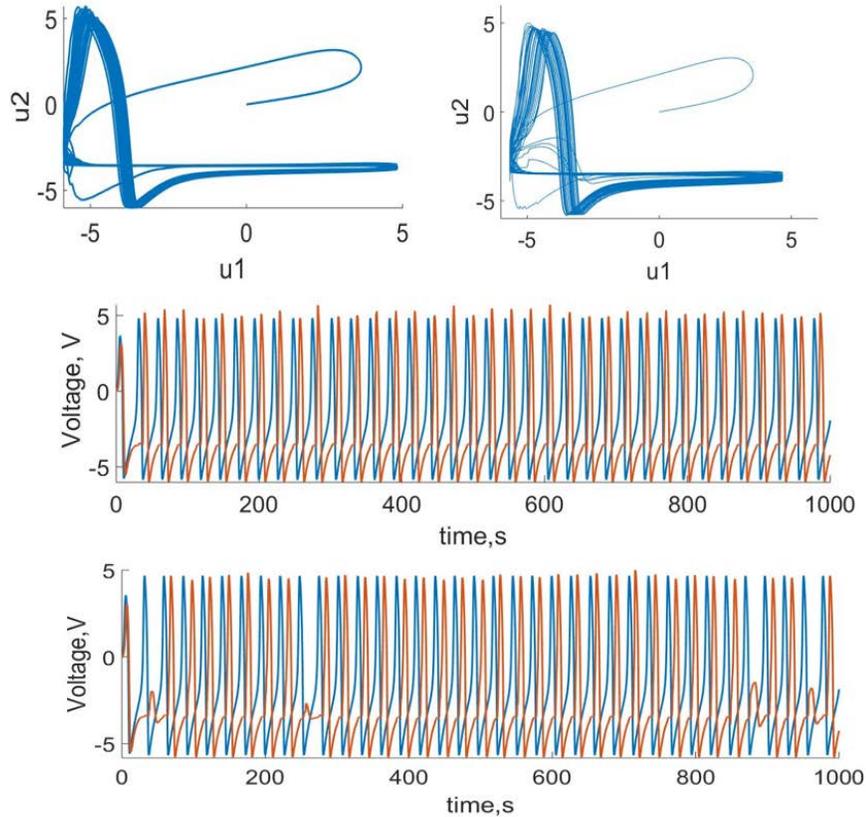

**Figure 3. Simulated synaptic coupling of neuron-like generators.**

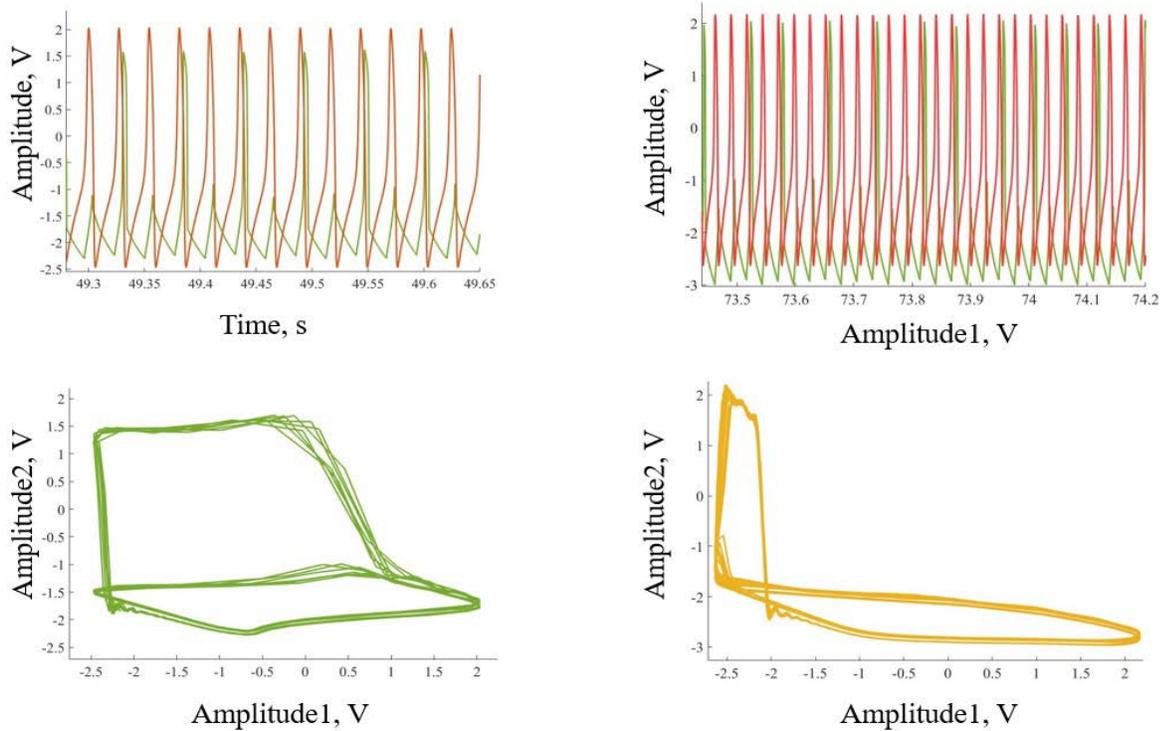

**Figure 4. Experimental results of switching between synchronization modes due to memristive dynamics.**

## CONCLUSIONS

In this paper, the main characteristics of the effect of synchronization of electronic neurons interacting via the memristive device have been shown and studied. The observed synchronization effects are associated with several central issues of neuroscience. It is generally accepted that synchronization is one of the most significant mechanisms of information processing by neurons in various areas of the brain, as well as communication between these areas. The synchronous generation of spikes by many neurons underlies important cognitive functions such as memory, attention, learning, perception. We have shown that this important behavior can be simulated by using a simple memristor model. However, to describe the rich experimentally observed synchronization modes, higher-order memristor models are required.

We believe that due to its relative compactness and high sensitivity, the experimental implementation of this memristive synapse is very promising for designing a large neuron network for biorobotic and bioengineering applications (Parvizi-Fard, 2021).

## CONFLICT OF INTEREST

The authors declare that the research was conducted in the absence of any commercial or financial relationships that could be construed as a potential conflict of interest.

## AUTHOR CONTRIBUTIONS

All authors gave substantial contribution to the development of this work equally, drafting and revising it critically; furthermore, all Authors approved its final version for publication.

**FUNDING**

The study was supported by the Lobachevsky University Competitiveness Program in the frame of the 5-100 Russian Academic Excellence Project.

**REFERENCES**

Adamatzky, A., and Chua, L. O. (2014). *Memristor Networks*, eds. A. Adamatzky and L. Chua. Cham: Springer International Publishing doi:10.1007/978-3-319-02630-5.

Adamchik, D. A., Matrosov, V. V., Semyanov, A. V., and Kazantsev, V. B. (2015). Model of self-oscillations in a neuron generator under the action of an active medium. *JETP Lett.* 102, 624–627. doi:10.1134/S0021364015210031.

Alombah, N. H., Fotsin, H., and Romanic, K. (2017). Coexistence of Multiple Attractors, Metastable Chaos and Bursting Oscillations in a Multiscroll Memristive Chaotic Circuit. *Int. J. Bifurc. Chaos* 27, 1750067. doi:10.1142/S0218127417500675.

Bill, J., and Legenstein, R. (2014). A compound memristive synapse model for statistical learning through STDP in spiking neural networks. *Front. Neurosci.* 8, 1–18. doi:10.3389/fnins.2014.00412.

Binczak, S., Jacquir, S., Bilbault, J.-M., Kazantsev, V. B., and Nekorkin, V. I. (2006). Experimental study of electrical FitzHugh–Nagumo neurons with modified excitability. *Neural Networks* 19, 684–693. doi:10.1016/j.neunet.2005.07.011.

Boccaletti, S., Pisarchik, A. N., del Genio, C. I., and Amann, A. (2018). *Synchronization: From Coupled Systems to Complex Networks*. Cambridge University Press doi:10.1017/9781107297111.

Choi, S., Tan, S. H., Li, Z., Kim, Y., Choi, C., Chen, P. Y., et al. (2018). SiGe epitaxial memory for neuromorphic computing with reproducible high performance based on engineered dislocations. *Nat. Mater.* 17, 335–340. doi:10.1038/s41563-017-0001-5.

Emelyanov, A. V., Nikiruy, K. E., Demin, A. V., Rylkov, V. V., Belov, A. I., Korolev, D. S., et al. (2019). Yttria-stabilized zirconia cross-point memristive devices for neuromorphic applications. *Microelectron. Eng.* 215. doi:10.1016/j.mee.2019.110988.

Gambuzza, L. V., Frasca, M., Fortuna, L., Ntinas, V., Vourkas, I., and Sirakoulis, G. C. (2017). Memristor Crossbar for Adaptive Synchronization. *IEEE Trans. Circuits Syst. I Regul. Pap.* 64, 2124–2133. doi:10.1109/TCSI.2017.2692519.

García-Vellisca, M. A., Jaimes-Reátegui, R., and Pisarchik, A. N. (2017). Chaos in Neural Oscillators Induced by Unidirectional Electrical Coupling. *Math. Model. Nat. Phenom.* 12, 43–52. doi:10.1051/mmnp/201712405.

Ge, R., Wu, X., Kim, M., Shi, J., Sonde, S., Tao, L., et al. (2017). Atomristor: Nonvolatile Resistance Switching in Atomic Sheets of Transition Metal Dichalcogenides. *Nano Lett.* 18, 434–441. doi:10.1021/acs.nanolett.7b04342.

Gerasimova, S. A., Gelikonov, G. V., Pisarchik, A. N., and Kazantsev, V. B. (2015). Synchronization

of optically coupled neural-like oscillators. *J. Commun. Technol. Electron.* 60, 900–903. doi:10.1134/s1064226915070062.

Gerasimova, S. A., Mikhaylov, A. N., Belov, A. I., Korolev, D. S., Gorshkov, O. N., and Kazantsev, V. B. (2017). Simulation of synaptic coupling of neuron-generators via a memristive device. *Tech. Phys.* 62, 1259–1265. doi:10.1134/s1063784217080102.

Gerasimova, S. A., Mikhaylov, A.N., Belov, A.I., Korolev, D.S., Guseinov, D.V., Lebedeva, A.V., Gorshkov, O.N. and Kazantsev, V.B. (2018). Design of memristive interface between electronic neurons. *AIP Conference Proceedings.* 1, 1959. doi:10.1063/1.5034744

Guseinov, D. V., Tetelbaum, D. I., Mikhaylov, A. N., Belov, A. I., Shenina, M. E., Korolev, D. S., et al. (2017). Filamentary model of bipolar resistive switching in capacitor-like memristive nanostructures on the basis of yttria-stabilised zirconia. *Int. J. Nanotechnol.* 14, 604. doi:10.1504/IJNT.2017.083436.

Guseinov, D. V., Matyushkin, I. V., Chernyaev, N. V., Mikhaylov, A. N., and Pershin, Y. V. (2021). Capacitive effects can make memristors chaotic. *Chaos, Solitons & Fractals* 144, 110699. doi:10.1016/j.chaos.2021.110699

Horch, K. W., and Dhillon, G. S. (2004). *Neuroprosthetics Theory and Practice*.

Ignatov, M., Ziegler, M., Hansen, M., Petraru, A., and Kohlstedt, H. (2015). A memristive spiking neuron with firing rate coding. *Front. Neurosci.* 9, 1–9. doi:10.3389/fnins.2015.00376.

Indiveri, G., Linares-Barranco, B., Hamilton, T. J., Schaik, A. van, Etienne-Cummings, R., Delbruck, T., et al. (2011). Neuromorphic Silicon Neuron Circuits. *Front. Neurosci.* 5. doi:10.3389/fnins.2011.00073.

Korotkov, A.G., Kazakov, A.O., Levanova, T.A. and Osipov, G.V. (2019). The dynamics of ensemble of neuron-like elements with excitatory couplings. *Communications in nonlinear science and numerical simulation*, 71, 38-49. doi:10.1016/j.cnsns.2018.10.023

Kuzum, D., Yu, S., and Philip Wong, H. S. (2013). Synaptic electronics: Materials, devices and applications. *Nanotechnology* 24. doi:10.1088/0957-4484/24/38/382001.

Matrosov, V. V., and Kazantsev, V. B. (2011). Bifurcation mechanisms of regular and chaotic network signaling in brain astrocytes. *Chaos An Interdiscip. J. Nonlinear Sci.* 21, 023103. doi:10.1063/1.3574031.

Matrosov, V. V., Mishchenko, M. A., and Shalfeev, V. D. (2013). Neuron-like dynamics of a phase-locked loop. *Eur. Phys. J. Spec. Top.* 222, 2399–2405. doi:10.1140/epjst/e2013-02024-9.

Matveyev, Y., Egorov, K., Markeev, A., and Zenkevich, A. (2015). Resistive switching and synaptic properties of fully atomic layer deposition grown TiN/HfO$_2$/TiN devices. *J. Appl. Phys.* 117, 044901. doi:10.1063/1.4905792.

Mikhaylov, A., Pimashkin, A., Pigareva, Y., Gerasimova, S., Gryaznov, E., Shchanikov, S., et al. (2020). Neurohybrid Memristive CMOS-Integrated Systems for Biosensors and Neuroprosthetics. *Front. Neurosci.* 14, 358. doi:10.3389/fnins.2020.00358.

Mishchenko, M. A., Bolshakov, D. I., and Matrosov, V. V. (2017). Instrumental implementation of a neuronlike generator with spiking and bursting dynamics based on a phase-locked loop. *Tech. Phys. Lett.* 43, 596–599. doi:10.1134/S1063785017070100.

Nair, M. V., Muller, L. K., and Indiveri, G. (2017). A differential memristive synapse circuit for on-line learning in neuromorphic computing systems. *Nano Futur.* 1, 035003. doi:10.1088/2399-1984/aa954a.

Parvizi-Fard, A., Amiri, M., Kumar, D., Iskarous, M. M. and Thakor, N. V. (2021) A functional spiking neuronal network for tactile sensing pathway to process edge orientation. *Scientific reports*, 11.1, 1-16. doi: 10.1038/s41598-020-80132-4.

Pisarchik, A. N., Jaimes-Reátegui, R., Sevilla-Escoboza, R., García-Lopez, J. H., and Kazantsev, V. B. (2011). Optical fiber synaptic sensor. *Opt. Lasers Eng.* 49, 736–742. doi:10.1016/j.optlaseng.2011.01.020.

Pisarchik, A., Sevilla-Escoboza, R., Jaimes-Reátegui, R., Huerta-Cuellar, G., García-Lopez, J., and Kazantsev, V. (2013). Experimental Implementation of a Biometric Laser Synaptic Sensor. *Sensors* 13, 17322–17331. doi:10.3390/s131217322.

Selyutskiy, Y. D. (2017). On auto-oscillations of a plate in flow. in *AIP Conference Proceedings*, 020139. doi:10.1063/1.4972731.

Shchapin, D. S. (2009). Dynamics of two neuronlike elements with inhibitory feedback. *J. Commun. Technol. Electron.* doi:10.1134/S1064226909020089.

Shi, Y., Fong, S., Wong, H.-S. P., and Kuzum, D. (2017). "Synaptic Devices Based on Phase-Change Memory," in *Neuro-inspired Computing Using Resistive Synaptic Devices* doi:10.1007/978-3-319-54313-0_2.

Strukov, D. B. (2018). Tightening grip. Nat. Mater. 17, 293–295. doi:10.1038/s41563-018-0020-x.

Thomas, A. (2013). Memristor-based neural networks. J. Phys. D. Appl. Phys. 46. doi:10.1088/0022-3727/46/9/093001.

Zhang, T., Yin, M., Lu, X., Cai, Y., Yang, Y., and Huang, R. (2017). Tolerance of intrinsic device variation in fuzzy restricted Boltzmann machine network based on memristive nano-synapses. Nano Futur. 1, 015003. doi:10.1088/2399-1984/aa678b